\documentclass[twocolumn,showpacs,preprintnumbers,amsmath,amssymb,prb]{revtex4}
\usepackage{graphicx}% Include figure files
\usepackage{dcolumn}% Align table columns on decimal point
\usepackage{bm}% bold math
\newcommand{\dir}{Figs}

\begin{document}

%\preprint{}

\title{Screening of hydrodynamic interactions for polyelectrolytes in salt solution}

% \altaffiliation[Also at ]{Physics Department, XYZ University.}
%\affiliation{%

\author{Jens Smiatek, Friederike Schmid}
%\homepage{http://www.physik.uni-bielefeld.de/theory/cm/}
\affiliation{
Condensed Matter Theory Group, Fakult\"at f\"ur Physik, Universit\"at Bielefeld
}%

\date{September 30, 2008}% It is always \today, today,

\begin{abstract}
We provide numerical evidence that hydrodynamic interactions are screened for 
charged polymers in salt solution on time scales below the Zimm time. At very short 
times, a crossover to hydrodynamic behavior is observed.
Our conclusions are drawn from extensive coarse-grained computer simulations 
of polyelectrolytes in explicit solvent and explicit salt, and discussed
in terms of analytical arguments based on the Debye-H\"uckel approximation.
\end{abstract}

\maketitle

Macromolecules of biological relevance such as DNA are often highly charged
polyelectrolytes. Under physiological conditions, these molecules are dissolved 
in salt solutions with Debye lengths of less than 1 nm. From the point of view
of statics, the electrostatic interactions are thus screened: Large DNA 
strands in physiological buffers have similar static properties than 
regular self-avoiding chains with short range interactions.
The dynamical properties are more complex. For neutral chains, solvent-mediated 
hydrodynamic interactions influence the mobility and the internal modes of
the molecules. The signature of these hydrodynamic interactions is
dynamical ``Zimm scaling''\cite{DE}. For example, the mobility 
$\mu$ and the diffusion constant $D$ of a self-avoiding Zimm chain scale 
as $\mu\! \sim \! D \sim \!N^{-\nu}$ with the chain length $N$, where 
$\nu = 0.588$ is the well-known Flory exponent. In contrast, the 
same quantities scale as $N^{-1}$ in a ``Rouse'' 
chain, where hydrodynamic interactions are absent or screened.

For polyelectrolytes like DNA, the interaction with the counterions
and salt ions complicates the situation. The polyelectrolyte chain is
surrounded by an oppositely charged ion cloud, which drags behind and 
generates additional friction \cite{BJ}. This ``relaxation
effect'' reduces the mobility, but the dominating behavior for large 
chain lengths is still predicted to be Zimm-like, and indeed, the
scaling exponents reported in experiments (e.g., $D \sim N^{-0.67}$ 
for double stranded DNA in Ref.~\onlinecite{SLS03}) are closer 
to the Zimm than to the Rouse exponent.

On the other hand, it has long been known that the {\em electrophoretic}
mobility in an applied electrical field scales like $\mu_e \sim N^{-1}$ 
for long chains, {\em i.e.}, it exhibits Rouse-like behavior 
\cite{Mann81}. The physical explanation for this effect 
is rather intuitive: The electrical field acts both on the 
polyelectrolyte and the surrounding ion cloud. The net momentum 
transferred to the solvent by all particles is thus zero, and 
the hydrodynamic interactions are screened as a result. 
One can derive a screened Oseen tensor, which describes 
hydrodynamic interactions between monomers in a homogeneous 
electrical field \cite{Mann81,BJ}.

Hence it appears that electrostatic and hydrodynamic screening are
closely related to each other, and that electrostatic screening may
entail hydrodynamic screening if the dynamics is driven by
electrical fields. Electrostatic forces drive electrophoresis, 
but also to some extent the internal motion of polyelectrolytes, 
since they dominate the nonbonded interactions between monomers. 
The question thus arises whether hydrodynamic screening can also 
be observed in the absence of external fields.

This question is addressed in the present letter. We consider a 
polyelectrolyte in salt solution, without external fields, and 
study the dynamics of internal motions in the chain. The dynamical 
scaling -- Zimm vs. Rouse -- gives us information on the influence of
hydrodynamic interactions. We use a generic coarse-grained model both
for the polyelectrolyte and the solvent. Recently, it has been shown that
such models can account quantitatively for the dynamics of real
polyelectrolytes in solution \cite{GBS08,FW08}. 

On general grounds, we already know that the dynamics on very large 
time scales, where chain diffusion dominates, must be Zimm-like: By virtue 
of the Einstein relation, the diffusion constant is related to the 
mobility of the chain in a sedimentation field, which acts only 
on the monomers and not on the ions. The interaction between monomers
and ions leads to a relaxation effect, as discussed above, but not
to hydrodynamic screening. Diffusion-dominated behavior sets in 
at the ``Zimm time'', {\em i.e.}, the time scale on which single monomers
follow the center of the mass motion of the whole chain. In the
present work, we thus focus on time scales below the Zimm time, where 
the dynamics is governed by the internal modes of the chain.

We study the dynamics of single bead-spring chains immersed in a 
solution of solvent particles and ions (salt ions and counterions) by 
dissipative particle dynamics (DPD) simulations \cite{EW}. Ions and 
monomers are hard-core particles with diameter $\sigma$, which interact with 
the potential $V_{\mbox{\tiny hc}} = 4 \epsilon ((r/\sigma)^{12}-(r/\sigma)^6)$
at $r < \sigma$. In addition, the ions and a fraction of the 
monomers carry single-valued charges $\pm e$ and interact {\em via}
Coulomb potentials. The monomers in the chain are connected by finitely 
extensible nonlinear elastic (FENE) springs with the bond potential
$V_{\mbox{\tiny bond}} = 1/2 \: k \: R^2 \ln(1-(r/R)^2)$.
The solvent particles provide the fluid background and have no 
conservative interactions. All particles interact with 
dissipative DPD forces, which have the usual form \cite{EW} 
${\bf F}_{ij}^{DPD} = {\bf F}_{ij}^{D} + {\bf F}_{ij}^{R}$
with a viscous contribution ${\bf F}_{ij}^{D} = - \gamma \omega(r_{_{ij}}) 
(\hat{r}_{_{ij}} \cdot {\bf v}_{_{ij}}) \hat{r}_{_{ij}}$ and a random force
${\bf F}_{_{ij}}^{R} = \sqrt{2 \gamma k_{_B} T \omega(r_{_{ij}})} \hat{r}_{_{ij}}
\chi_{_{ij}}$.
Here ${\bf v}_{_{ij}}$ is the velocity difference and ${\bf r}_{_{ij}}$
the distance between the two particles $i$ and $j$, 
$\hat{r}_{_{ij}} = {\bf r}_{_{ij}}/r_{_{ij}}$, $\omega(r) = 1-(r/\sigma)$ for
$r < \sigma$, and $\chi_{_{ij}}$ is a Gaussian distributed white noise 
with mean zero and variance one. All particles have the same mass $m$.
The natural units in our system are thus the length $\sigma$, the 
thermal energy $k_{_B} T$, and the time unit $\tau = \sqrt{m/k_{_B} T} \sigma$.

Specifically, we consider chains of length $N=50$ with charges on every
second monomer, in solutions with ion concentration $\rho_{_i} = 0.1 \sigma^{-3}$.
The other model parameters are $e = \sqrt{\sigma k_{_B} T}$, $\epsilon = k_{_B} T$, 
$R = 1.5 \sigma$, $k = 25 k_{_B} T/\sigma^2$, and 
$\gamma = 5 k_{_B} T \, \tau/\sigma^2$, and the
density of solvent particles is $\rho_{_s} = 3 \sigma^{-3}$. With
these parameters, the motion of the ions in solution is diffusive after 
an initial ballistic regime of length $\sim 10 \tau$.
The Debye length in the solution is $\lambda_{_D}=0.89 \sigma$,
the kinematic viscosity of the fluid is $\eta = 1.24 k_{_B} T \tau/\sigma^3$,
and the ions have the mobility $\zeta = 0.44 \sigma^2/\tau k_{_B} T$. 
The mean radius of {\em uncharged} chains of length $N=50$ is
$R_{_{g,u}} = 4.74 \sigma$, and they diffuse with the diffusion constant
$D{_u} = 0.021 \sigma^2/\tau$. In comparison, the charged chains are just
slightly more swollen, $R_{_{g,c}} = 5.35 \sigma$, but their diffusion constant
is significantly reduced due to the relaxation effect, 
$D{_c} = 0.013 \sigma^2/\tau$. 
The simulations were carried out in cubic simulation boxes with system 
size $L=25 \sigma$ and periodic boundary conditions, using the time step
$\Delta t = 0.01 \tau$. The run lengths were 4 million 
time steps after an equilibration time of 2 million steps. 

\begin{figure}[t]
\includegraphics[width=0.35 \textwidth]{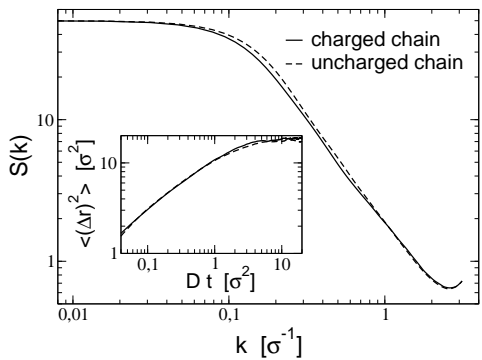}
\caption{Static structure factor $S(k)$ of the charged chain in salt solution 
  (solid line) compared to that of an uncharged chain (dashed line). The inset 
  shows the mean square displacement of the central monomer relative to the 
  center of mass of the chain $\langle (\Delta r)^2 \rangle$ as a function of 
  the rescalce time $Dt$ with the diffusion constant $D$.
         }
\label{fig:struc_stat}
\end{figure}

Fig.\ \ref{fig:struc_stat} compares the static structure factor for such 
a charged chain in salt solution to that of an uncharged chain with the same 
length. The charged chain is slightly stretched, otherwise the two structure 
factors are very similar. Algebraic scaling behavior $S(k) \propto k^{-\nu}$ 
is observed in the wavevector range $1/R_{_g}  < k < 1/a_{_0}$, where 
$a_{_0}\sim 1.7 \sigma$ is the distance between charged monomers in the chain. 
The value of $\nu$ is found to be $\nu = 0.67$ for the uncharged chain, and 
$\nu = 0.7$ for the charged chain. These values are larger than the universal 
asymptotic value, $\nu = 0.588$, due to the finite length of the chains.

The inset of Fig.\ \ref{fig:struc_stat} shows the mean square displacement 
of the central monomer in the reference frame of the chain,
$\langle (\Delta r)^2 \rangle\! =\! \langle \big({\bf r}_{_i}\!(t)\!-\!
{\bf R}_{_{cm}}\!(t)\!-\! {\bf r}_{_i}\!(0)\!+\!
{\bf R}_{_{cm}}\!(0) \big)^2 \rangle $, as a function of time for charged 
and uncharged chains. If the time is scaled with the diffusion constant $D$, 
the two curves lie almost on top of each other, indicating that internal modes 
are slowed down by the relaxation effect in the same way than the overall chain 
diffusion. The quantity $\langle (\Delta r)^2 \rangle$ reaches a 
plateau at the Zimm time $t_{_{\mbox{\tiny Zimm}}}$. From Fig.\ \ref{fig:struc_stat}, 
we infer $D t_{_{\mbox{\tiny Zimm}}}\!\! \sim \!4 \sigma^2$, in rough agreement with the
heuristic estimate $6 D t_{_{\mbox{\tiny Zimm}}}\!\! \sim \!R_{_g}^2$ \cite{AD99}.
This yields $t_{_{\mbox{\tiny Zimm}}}\!\! \sim \!300 \tau$ for the charged chain,
and $t_{_{\mbox{\tiny Zimm}}}\!\! \sim \!200 \tau$ for the uncharged chain. 

\begin{figure}[t]
\includegraphics[width=0.45 \textwidth]{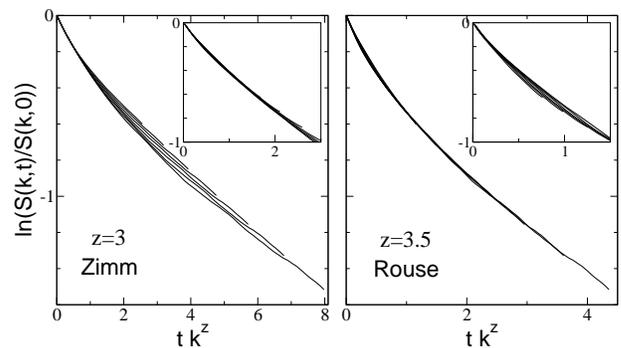}
\caption{Dynamic structure factor of the charged chain scaled with 
  Zimm (left) and Rouse (right) exponent $z$ for wavevectors in the
  range $0.2 < k \sigma < 0.3$ (large wavelengths) and times
  $t < t_{_{\mbox{\tiny Zimm}}}$. The inset shows the corresponding 
  data for the uncharged chain.
         }
\label{fig:struc_dyn1}
\end{figure}

To assess the dynamical properties of the chains, we study the dynamic
structure factor $S(k,t)$. In the dynamic scaling regime \cite{DE,DK93}, 
it should exhibit the universal scaling behavior $S(k,t) = S(k,0) f(t k^z)$ 
with the exponent $z=3$ for Zimm chains and $z =  2 + 1/\nu$ for Rouse chains. 
Hence the curves $S(k,t)/S(k,0)$ for different $k$ vectors should collapse
onto a single master curve $f(t k^z)$ when plotted against $t k^z$ with the correct
exponent $z$. Fig.\ \ref{fig:struc_dyn1} shows such scaling plots for
large wavelengths, $1/R_{_g} < k < 0.3 \sigma^{-1}$, using $\nu = 0.67$. 
Not surprisingly, uncharged chains exhibit Zimm scaling. For charged 
chains, however, the data collapse much better in the Rouse scaling plot 
than in the Zimm scaling plot: Below the Zimm time, the chain behaves as 
if hydrodynamic interactions were absent.

\begin{figure}[t]
\includegraphics[width=0.45 \textwidth]{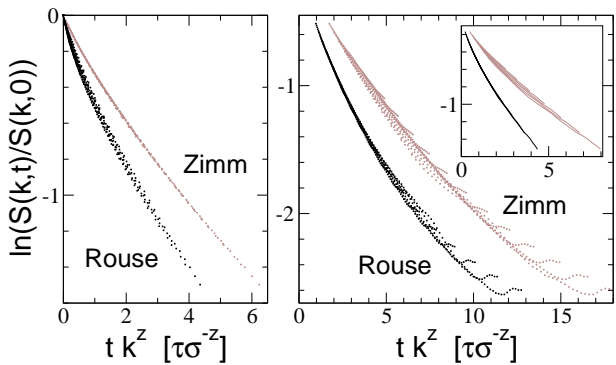}
\caption{Dynamic structure factor of the charged chain scaled with
  Zimm ($z=3$, light circles) and Rouse ($z=3.5$, black circles)
  for wavevectors $0.2 < k \sigma < 0.5$ 
  and times $t < t_{_0} = 55 \tau$ (left) and $t > t_{_0}$ (right).
  The inset on the right shows data for large wavelengths,
  $k < k_{_0} = 0.3/\sigma$ at 
  $t_{_0} < t < t_{_{\mbox{\tiny Zimm}}} = 300 \tau$, 
  which exhibit Rouse scaling as already demonstrated in 
  Fig.\ \protect\ref{fig:struc_dyn1}. The main graph shows the
  data for short wavelengths, $k > k_{_0} = 0.3/\sigma$ and
  times $t_0 < t < 150\tau$. 
         }
\label{fig:struc_dyn2}
\end{figure}

A closer inspection of Fig.\ \ref{fig:struc_dyn1} shows that this
is not yet the full story. At very early times, the data collapse in
the Zimm plot seems more convincing than in the Rouse plot. Indeed, we
can identify a second characteristic time $t_{_0} \sim 55 \tau$, below which the
data show Zimm scaling for a range of $k$-vectors that includes both
long and shorter length scales, $1/R_{_g} < k < 0.5 \sigma^{-1}$  
(Fig.\ \ref{fig:struc_dyn2}). For $t > t_{_0}$, the Zimm scaling breaks down
for {\em all} $k$. For $k < k_{_0} \sim 0.3 \sigma^{-2}$, one has a crossover 
to Rouse scaling. For $k > k_{_0}$, the data do not scale at all.

To summarize, we find that hydrodynamic interactions are screened on large 
length scales. However, the screening is `delayed' and not yet effective at 
early times $t < t_{_0} \approx 55 \tau$. A similar effect has been observed by 
Ahlrichs {\em et al.} in semidilute polymer melts \cite{AEB}. In this case,
the initial decay of $S(k,t)$ was found to be governed by an unscreened diffusion
tensor. Screening sets in as soon as chains interact with each other, {\em i.e.},
mesh blobs have moved their own size. Beyond that time, chain parts cannot 
follow the flow and act as immobile obstacles that produce Darcy-type friction. 

In our case, the phenomenon is similar, but the underlying physics is clearly 
different. The large reduction of the diffusion constant in the presence of 
charges ($D{_c} < D{_u}$) suggests that the chain dynamics is mainly driven by 
electrostatics, like in electrophoresis, where hydrodynamic screening is also 
observed. However, the derivation of the screened Oseen tensor in electrophoresis 
relies crucially on the fact that the monomer and the surrounding ion cloud 
are subject to a {\em homogeneous} electrical field. Intrachain interactions produce 
inhomogeneous fields, thus the theory cannot be applied. Moreover, we need to explain 
the screening delay, {\em i.e.}, the Zimm-Rouse crossover observed in our simulations.

\begin{figure}[t]
\includegraphics[width=0.48 \textwidth]{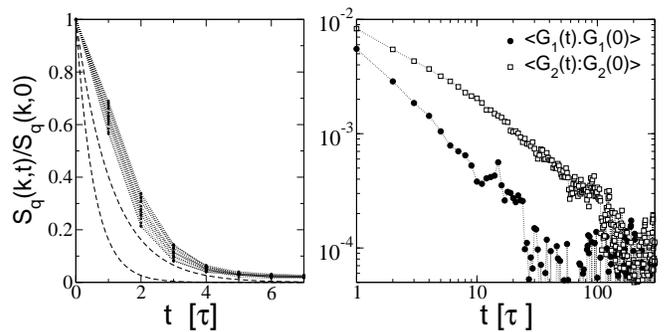}
\caption{Illustration of time scales governing the charged system.
  Left: dynamic charge structure factor in the range $k \lambda_{_D} < 1$. 
  Dashed lines show theoretically predicted
  exponential decay $\exp(-t/\tau(k))$ (see text) for $k = 0$ (upper curve) and 
  $k = 1/\lambda_{_D}$ (lower curve).
  Right: Autocorrelation function of ${\bf G}_1(t)$ and ${\bf G}_2(t)$ (see text 
  for definitions). Here $\cdot$ denotes dot product and $:$ double-dot
  product (double contraction).
         }
\label{fig:timescales}
\end{figure}

To analyze the problem in more detail, we begin with investigating the 
time scales governing the dynamics of charge distributions in our system. 
To this end, we calculate the dynamic ''charge structure factor'' 
$S_{q}(k,t) = \langle |I_q({\bf k},t) I_q(-{\bf k},0)|\rangle$ with
$I_q = \sum_{i} q_{_i} e^{i {\bf k} \cdot  {\bf R}_{_i}}$, where the sum $i$ 
runs over all charged particles and $q_{_i}$ denotes their charge. The result is 
shown in Fig.~\ref{fig:timescales} (left). In the simplest linearized
dynamical mean-field theory, the charge distribution decays exponentially 
towards the Debye-H\"uckel distribution with the characteristic decay time 
$\tau_{_D}(k) = 1/\zeta k_{_B} T (k^2 + \lambda_{_D}^{-2})$. 
In the simulations, the decay of $S_{q}(k,t)$ is slightly slower and depends less
on $k$ -- presumably due to the fact that the motion of the ions is not yet 
diffusive on the time scale $\tau_{_D}$.  Nevertheless, the charge distribution 
has clearly relaxed on the time scale $t_{_0}$ of the Zimm-Rouse crossover. On 
this time scale, the ions follow the chain adiabatically.

We thus proceed to calculate the velocity field ${\bf v}$ generated
by the chain in the Debye-H\"uckel and Stokes approximation. For simplicity, 
we neglect the finite diameter of monomers and ions. The chain is characterized
by the positions ${\bf R}_{_{\alpha}}$ of the monomers $\alpha$ and their
central charge $q_{_{\alpha}}$. Each monomer charge generates a charge distribution 
$\rho({\bf r})$, which in turn generates an electrostatic field ${\bf E}({\bf r})$ 
and imparts a force field 
${\bf f}_{_{\mbox{\tiny el.}}} ({\bf r}) = \rho({\bf r}) {\bf E}({\bf r})$ 
to the fluid. In addition, non-electrostatic interactions ({\em e.g.}, 
bond interactions) between monomers $\alpha$ and $\beta$ generate forces 
$f_{_{\alpha \beta}}(r_{_{\alpha \beta}}) \hat{r}_{_{\alpha \beta}}$ 
at ${\bf r}={\bf R}_{_{\alpha}}$, which are also transmitted to the fluid. Here 
${\bf r}_{_{\alpha \beta}} = {\bf R}_{_{\beta}}-{\bf R}_{_{\alpha}}$
and $\hat{r}={\bf r}/r$. 

We consider the limit $k \ll 1/\lambda_{_D}$, {\em i.e.}, we focus on the 
length scale regime well above the Debye-H\"uckel screening length. After
working out the theory, we obtain the velocity field
${\bf v}({\bf k}) = \sum_{\beta} \exp(i {\bf k} \cdot {\bf R}_{_{\beta}}) \:
{\bf v}^{(\beta)}({\bf k})$,
\begin{equation}
\label{eq:vk}
{\bf v}^{(\beta)}({\bf k}) = \frac{1}{\eta k^2} \!
\sum_{\gamma}\!  \Big \{
f_{_{\beta \gamma}}
\!+\!
\frac{
q_{_{\beta}} q_{_{\gamma}}}{\lambda_{_D}^2} 
g\big(\frac{r_{_{\beta \gamma}}}{\lambda_{_D}}\big)
e^{i {\bf k} \cdot {\bf r}_{_{\beta \gamma}}}  
\Big\} 
({\bf 1} - \hat{k}\hat{k}) \cdot \hat{r}_{_{\beta \gamma}}
\end{equation}
with $g(x) = e^{-x} \: (x^2/2 - x - 1)/x^2$ and $\hat{k} = {\bf k}/k$,
where $\hat{k} \hat{k}$ denotes the tensor product. A derivation of this 
expression shall be given elsewhere.
From Eq.~(\ref{eq:vk}), one readily calculates the flow velocity
at the position of a monomer $\alpha$, ${\bf v}({\bf R}_{_{\alpha}}) = 
\sum_{\beta} {\bf v}^{(\beta)}({\bf r}_{_{\beta \alpha}})$.
The contributions ${\bf v}^{(\beta)}$ are conveniently expanded in multipoles.
Here we quote only the monopole and dipole term
\begin{eqnarray}
\label{eq:vr1}
{\bf v}^{(\beta)}_m({\bf r})\! &=&\! \frac{1}{8 \pi \eta r} \!
\sum_{\gamma}  \Big \{
f_{_{\beta \gamma}}
\!+\!
\frac{
q_{_{\beta}} q_{_{\gamma}}}{\lambda_{_D}^2} 
g\big(\frac{r_{_{\beta \gamma}}}{\lambda_{_D}}\big)
\Big\} 
({\bf 1} + \hat{r} \hat{r}) \cdot \hat{r}_{_{\beta \gamma}} \\
\label{eq:vr2}
{\bf v}^{(\beta)}_d({\bf r})\! &=&\! \frac{1}{8 \pi \eta}
\frac{{\bf r}}{r^3} \!
\sum_{\gamma}  
\frac{
q_{_{\beta}} q_{_{\gamma}}}{\lambda_{_D}^2} \: 
g\big(\frac{r_{_{\beta \gamma}}}{\lambda_{_D}}\big) \: 
{\bf r}_{_{\beta \gamma}} \cdot
(3 \hat{r} \hat{r} \!-\! {\bf 1}) \cdot \hat{r}_{_{\beta \gamma}}.
\end{eqnarray}
The non-electrostatic interactions $f_{_{\beta \alpha}}$ between 
monomers only contribute to the monopole term, Eq.~(\ref{eq:vr1}). 
Inspecting the latter, one notices that it vanishes for
locally straight chains, because contributions of vectors 
${\bf r}_{_{\beta \gamma}}$ are then balanced by contributions of vectors 
${\bf r}_{_{\beta \gamma'}} = - {\bf r}_{_{\beta \gamma}}$. For symmetry 
reasons, the time-averaged conformations of the chain have this property.
Therefore, one might speculate that the monopole becomes small compared
to the higher order multipoles on sufficiently large time scales. On the other 
hand, it is clear from Eq.~(\ref{eq:vk}) that {\em all} multipoles must vanish
on time scales where the vectors ${\bf r}_{_{\beta \gamma}}$ are
distributed isotropically. We thus need to compare the characteristic decay 
times for the monopole and the higher order multipoles quantitatively.
To this end, we have evaluated the autocorrelation function of the vector 
${\bf G}_{_1}(t) = \sum_{\gamma} q_{_{\beta}} q_{_{\gamma}} 
g(r_{_{\beta \gamma}}/\lambda_{_D}) \hat{r}_{_{\beta \gamma}}$, 
which characterizes the monopole term, and that of the traceless part 
${\bf G}_{_2}(t)$ of the tensor $\sum_{\gamma} q_{_{\beta}} q_{_{\gamma}} 
g(r_{_{\beta \gamma}}/\lambda_{_D}) {\bf r}_{_{\beta \gamma}} 
\hat{r}_{_{\beta \gamma}}$, which characterizes the dipole term. Here
$\beta$ was chosen to be the central monomer in the chain. The results are 
shown in Fig.~\ref{fig:timescales} (right). For ${\bf G}_1$, the decay is much 
faster than that for ${\bf G}_2$. By the time $t_{_0} (55 \tau)$, the function 
$\langle {bf G}_1(t) \cdot {\bf G}_1(0) \rangle$ has dropped by two orders of 
magnitude. This suggests that the dynamics of the chain is indeed governed 
by the higher order multipole interactions at later times. 
In the Stokes approximation, the latter vanish too, but much 
slower than the multipole term. If one includes nonlinear hydrodynamics, 
additional interactions come into play that presumably do not vanish at all.
For example, the inhomogeneous velocity field created by the monomer
$\beta$ in the vicinity of monomer $\alpha$ convects the ion cloud surrounding
$\alpha$. This contributes an effective additional force field in the
vicinity of ${\bf R}_{_{\alpha}}$, which in turn retroacts on $\beta$, 
leading to an induced hydrodynamic interaction that remains finite at all times.

Based on these considerations, we rationalize our simulation results as follows:
On short time scales, the monopole term, Eq.~(\ref{eq:vr1}) dominates, hydrodynamic 
interactions are effective and lead to Zimm scaling. On longer time scales, 
time-averaged chains become locally straight and higher order terms take over.
Zimm scaling is then replaced by Rouse scaling \cite{fn1}.
The relevant characteristic {\em length} scale for the process of local
chain straightening is twice the maximal distance between interacting monomers  --
in our case, roughly $\lambda_{_0} \sim 3.4 \sigma$, since the mean distance between 
charged monomers in the chain is roughly $a_0=1.7 \sigma$ and $\lambda_{_D} < a_0 $.
This corresponds well with the observation that the Rouse scaling breaks down 
for large wave vectors with $k < k_0 \sim 1/\lambda_0$. It also explains why 
the crossover time $t_0$ from Zimm to Rouse behavior does not seem to depend on the 
wavevektor $k$.

To summarize, we have studied the dynamical behavior of polyelectrolytes
in salt solution on time scales below the Zimm time. At very short times, 
they behave like regular self-avoiding chains that are subject to 
hydrodynamic interactions. In an intermediate time scale range between a 
microscopic crossover time and the Zimm time, however, the chain behaves
like a Rouse chain, {\em i.e.}, hydrodynamic interactions are effectively
screened. These findings should have implications for the interpretation 
of various processes of biological and nanotechnological interest. Many 
dynamical processes where polyelectrolytes are involved, such as DNA 
translocation through pores or RNA folding etc. are driven by internal 
chain modes. Our results indicate that the influence of hydrodynamics on 
these processes might be much smaller than is commonly assumed.  
Theoretical models that neglect hydrodynamics may be closer to 
reality than more refined models that include hydrodynamics.

We thank Burkhard D\"unweg and Ulf Schiller for enlightening discussions. 
This work was funded by the VW foundation. The simulations were carried 
out on supercomputers at the HLRS in Stuttgart, the NIC in J\"ulich, 
and the PC${}^2$ cluster in Paderborn, using the freely available software 
package ESPResSo \cite{Espresso}.

\bibliography{paper}% Produces the bibliography via BibTeX.

\begin{thebibliography}{12}
\bibitem[{fn1()}]{fn1}
\bibinfo{note}{Rouse scaling is expected for hydrodynamic interactions decaying
  faster than $r^{-1/\nu}$ \protect\cite{DE}}.
\end{thebibliography}

\end{document}